# Self-organized atomic nanowires of noble metals on Ge(001): atomic structure and electronic properties

J Schäfer[1], S Meyer[1], C Blumenstein[1], K Roensch[1], R Claessen[1],
S Mietke[2], M Klinke[2], T Podlich[2], R Matzdorf[2]
A A Stekolnikov[3], S Sauer[3] and F Bechstedt[3]

[1]*Physikalisches Institut, Universität Würzburg, 97074 Würzburg, Germany*

[2]*Institut für Physik, Universität Kassel, 34132 Kassel, Germany*

[3]*Institut für Festkörpertheorie und -optik, Universität Jena, 07743 Jena, Germany*

Email: joerg.schaefer@physik.uni-wuerzburg.de

**Abstract.** Atomic structures of quasi-one-dimensional (1D) character can be grown on semiconductor substrates by metal adsorption. Significant progress concerning study of their 1D character has been achieved recently by condensing noble metal atoms on the Ge(001) surface. In particular, Pt and Au yield high quality reconstructions with low defect densities. We reported on the self-organized growth and the long-range order achieved, and present data from scanning tunneling microscopy (STM) on the structural components. For Pt/Ge(001), we find hot substrate growth is the preferred method for self-organization. Despite various dimerized bonds, these atomic wires exhibit metallic conduction at room temperature, as documented by low-bias STM. For the recently discovered Au/Ge(001) nanowires, we have developed a deposition technique that allows complete substrate coverage. The Au nanowires are extremely well separated spatially, exhibit a continuous 1D charge density, and are of solid metallic conductance. In this review we present structural details for both types of nanowires, and discuss similarities and differences. A perspective is given for their potential to host a one-dimensional electron system. The ability to condense different noble metal nanowires demonstrates how atomic control of the structure affects the electronic properties.

**Contents:**

## 1. Introduction: Atomic nanowires

*1.1 Fundamentals*

    There are several central issues why today's research takes such an enormous interest in the field of atomic-scale nanostructures. For one, obviously the thrust is to further miniaturize electronic circuitry, and to realize electronic switching functions on an ever shrinking length scale. Secondly, the study of atomic processes on surfaces is of eminent importance to understand atomic bonding and molecular interactions, which does intimately relate to self-assembly of nanoscale structures and molecular aggregates, as well as to catalytic processes. Thirdly, and probably most importantly, in reducing the size to the single-atom scale, and specifically by constructing two-dimensional (2D) and even quasi-one-dimensional (1D) structures, the physics of the electronic charge carriers will change dramatically. In particular, one is used to a description of the electron states within the Fermi liquid picture. Also, one is familiar with treatments of phase transitions within the mean-field theory. The latter, e.g., is used successfully to describe the conventional superconducting ground state of metals within the BCS formalism. However, in turning to material systems that are essentially 1D in nature, there is indication that these scenarios do no longer apply. In this report, we pay particular attention to such effects, and we review in detail how self-organized nanowires form on the atomic scale, including latest results on new nanowire architectures realized on (001) semiconductor surfaces, and to which extent they yield a particularly close approach to a 1D system.

    Concerning the physics in solids with nearly 1D properties, they let us expect rather strong interactions of electrons and lattice, owing to a reduction of electrostatic screening. Specifically, the 1D nature of the electron system can be responsible for the occurrence of a charge density wave (CDW), implying a metal-insulator transition for the electron band concerned [1]. Moreover, quasi-1D systems

in the metallic phase offer an opportunity to study unusual physics that results from the predicted breakdown of the Fermi liquid picture [2]. The physics of electrons in low dimensions is dramatically different from that encountered in conventional three-dimensional bulk metals, and solid state theory predicts exotic many-body scenarios. Most prominent for 1D systems is the emergence of the so-called "Luttinger liquid", which results from a decoupling of the spin and charge degrees of freedom [2]. Requirements for experimental realizations have to take into account the fragility of the 1D regime, which can be affected by phonons, and may eventually be destroyed by coupling to the other dimensions [3]. Indication for spin-charge separation has been found in carbon nanotubes [4], lithographic semiconductor channels [5] and linearly bonded crystals [6].

As a different and still growing effort to achieve 1D electronic systems, realizations in chains of metal adsorbates on semiconductor surfaces have been studied in the last years. They allow notably good control of the structural parameters. On the surface of semiconductors such as silicon and germanium, 1D reconstructions, so-called *atomic nanowires*, have been identified. Such 1D alignment of adatoms in the reconstruction, even if formed by metal atoms, needs not necessarily to be metallic, as this depends on the degree of electron band filling of the final bonding structure, including both metal and semiconductor atoms. Some of these nanowires are indeed found to exhibit metallic character, e.g., those formed by In atoms on Si(111) [7,8,9] or Au atoms on Si(553) [10,11], and on Si(557) [12], which have been studied extensively in the last years. A signature that the quasi-1D regime is realized is the occurrence of a CDW, which can significantly alter the electronic states [13]. This has been seen for both types of these nanowires.

The In (4×1) chains on Si(111) were one of the first examples, where a CDW was demonstrated in a surface 1D system [9,14]. They are self-organized and do not require any further aid for alignment. The In chains undergo a phase transition upon cooling into a (4×2) phase, which is identified as a Peierls instability. Recent structural models discuss also an additional (8×2) phase as the lowest temperature state, which does include minor additional structural rearrangement [15]. Nonetheless, from angle-resolved photoemission it has been shown that a CDW nesting mechanism exists for driving a (4×2) phase. In Fermi surface data it emerges that one of three electron sheets contains piecewise parallel sections that can host a nesting vector $q = \frac{1}{2} G$. Thus spanning exactly half the Brillouin zone, this nicely explains the corresponding period doubling observed in real space [9,16]. This situation with a half-filled band thus represents the simplest case of a Peierls distortion. Moreover, it has even been observed that the In-based CDW nanowires exhibit precursor fluctuations above the transition temperature [9]. Recently, surface optical spectroscopy in the infrared range has successfully been performed *in situ* on the In/Si(111) chains [17], including the phase transition into the low-temperature phase, providing additional input for structural models.

The use of stepped, high-index templates for self-organized nanowires then represented a further step in the study of 1D chains. Key representatives are the Au chains on Si(557) and Si(553) substrates [12,18,19]. While on a plain substrate a Au (5×2) reconstruction has long since been known which is insulating, the picture changes drastically on the high-index planes of silicon. Here, the substrate becomes part of the nanowire structure, and the Au decoration serves to stabilize a certain step structure. In particular, the Au-stabilized Si terraces are metallic, and according to Fermi surface data show quasi-1D character. Upon cooling, both Au/Si(557) and Au/Si(553) show formation of a CDW. It is noteworthy, however, that these phase transitions occur soon below room temperature, i.e. the CDW condenses relatively easily [10]. On Si(557), also Pb can be condensed in long-range ordered form, and a rather specific phase transition is observed which suppresses the conductance perpendicular to the chains at low temperature [20,21].

Occurrence of a CDW phase transition generally depends on the degree of electron-lattice coupling, lattice stiffness, and the dimensionality of the electron system. In particular, wave function overlap with neighboring chains and into the supporting layer is a major influential factor. One should note that quantum theory does no allow condensation of a CDW at finite temperature in perfect 1D, as it is suppressed by fluctuations down to zero temperature [1]. Instead, a slight coupling to higher dimensions is needed to achieve CDW condensation ("quasi-1D regime"). This consideration is of funda-mental importance with respect to the two competing regimes of either a Fermi liquid (which is no longer valid if close to 1D) or a Luttinger liquid (when strictly 1D), respectively [2,3]. In other words, the CDW existing in a Fermi liquid must be suppressed in order to allow Luttinger liquid physics to occur. Regarding the architecture of these previously reported chains, one finds a relative ease of CDW condensation, which may indicate that they are structurally not perfectly separated. On high-index Si, the Au atoms are even submerged into the terraces [22]. It was thought that Au/Si(557) exhibits spin-charge separation as predicted for the strict 1D case [12], yet in subsequent work this claim has not been substantiated [18,19], and simple band structure was found to explain the observations. Hence the search for better defined 1D systems is ongoing.

*1.2 Systems on the Ge(001) surface*

In addressing alternative substrates or adsorbates, it is striking that the (001) face of the tetrahedral semiconductors has not been systematically explored in the past. Only very recently, nanowire formation on the (001) surface has received heightened attention. On Si(001), nanowires have been reported for rare earth adatoms, e.g. Gd [23,24], which are believed to form silicide structures that are in fact rather large concerning their diameter. Self-organized formation of silicide nanowires on Si(001) can also be induced by Dy [25]. Interestingly, their electronic character here depends on the substrate used. While on Si(001) these broad metallic nanowires exhibit a quasi-1D band structure [26], the corresponding structures grown on Si(557) will induce Si(111) facets and exhibit a two-dimensional band structure [27].

Formation of noble metal nanowires was reported for the Ge(001) surface, using Pt adatoms [28,29]. Pt atoms induce narrow nanowires on Ge(001), see Figure 1 (discussed in detail below), which exhibit sharply localized states [30]. Density functional theory relates this to Pt d-orbitals, which lead to a marginal tunneling conductivity at room temperature [31]. Gold atoms also grow in self-organized manner, and linear reconstructions can be achieved [32]. A local (4×2) unit cell was reported for excessive coverage of 1.5 monolayers (ML), with a high density of defects. Long-range order in these two noble metal reconstructions is achieved upon deposition of a submonolayer coverage of Pt or Au, at either elevated temperature or with subsequent annealing. The formation of nanowires by In atoms [33] on Ge(001) has also been reported.

Until today, however, the electronic properties of these systems on Ge(001) have remained largely unclear. In particular, the Pt nanowire system was reported to be in a dimer-distorted phase that could be indicative of a CDW, yet no evidence for a phase transition has been found in comparing scanning tunneling microscopy (STM) data at 300 K and 77 K [29]. A rather low conductivity at zero bias is reported in [28,29]. Thus a detailed analysis of the electron states close to the Fermi level is highly desirable. For Au, potential metallicity remained unknown [32].

In this report we review new STM studies of Pt and Au induced nanowires on the Ge(001) surface where particular attention is paid to the low-energy properties. For Pt nanowires, indications for CDW behavior, such as a dimerization of the metal adatoms, are investigated. STM images at high bias reveal dimers both along and sideways of the chains, consistent with structural support elements. Surprisingly, for low voltages corresponding to states near the Fermi level $E_F$, the dimerization is virtually lifted on the ridge of the nanowires. This observation relies on using adequately low tunneling currents. Spectroscopy data corroborate a finite conductivity. It thus emerges that the electron states near $E_F$ are essentially decoupled from the dimerized structure of the nanowire embankment.

We also report on Au-induced chains assembled on Ge(001) by low-coverage growth in novel c(8×2) long-range order. Analyzed by scanning tunneling microscopy (STM), metallic charge is spread out in 1D direction, unaffected by substrate bonds. The charge on the nanowire resides within single atom dimensions. Photoemission identifies a corresponding Au-induced metallic band. This renders the c(8×2) Au/Ge(001) chains a 1D model system with unexcelled confinement.

## 2. Experimental considerations

Experimentally, Ge substrates (n-doped, resistivity < 1 Ωcm) need to be prepared to high cleanliness, as carbon-based impurities represent notorious obstacles to formation of long-range order. This relates to the strong bonding between carbon and germanium, so that highest standards of ultra-high vacuum (UHV) processing are mandatory. Good results for the Ge(001) surface can be achieved by repeated cycles of Ar sputtering (1 keV) and annealing to 800 °C. This procedures results in large flat terraces of low defect density.

As an alternative pathway, Ge(001) substrates may be etched chemically [34], with the final step such that an intentional oxide is left behind [35]. The advantage of this procedure is that the substrate simply needs to be flashed in UHV in order to desorb the protective oxide. We have tested both of these two procedures and find that they yield equally good results. While the outcome of the sputter procedure relies on sufficient annealing to remove the induced defects, the chemical procedure is very sensitive to high cleanliness standards in the chemicals used. Both methods produce, after annealing or flashing, respectively, excellent long-range order of the plain Ge(001) surface. This is verified by low-energy electron diffraction (LEED), showing the coexistence of p(2×1) and c(4×2) order as is characteristic of the Ge(001) surface at room temperature [36,37].

For noble metal evaporation, an electron beam evaporator has been proven to be a source that can be easily controlled. In the case of Pt, the sample needs to receive a post-deposition anneal at ~ 600 – 700 °C to allow formation of the well-ordered nanowire reconstruction, unless a hot deposition is used (as described below in 3.1). It is generally assumed that the reconstruction is based on a ¼ ML coverage of Pt, as discussed in context of the structural model below. If using gold as the metal adatom, it was evaporated onto the sample kept at 500 °C, thereby favoring ordered c(8×2) nanowire growth. The optimum Au coverage determined with a quartz crystal monitor was slightly above 0.5 ML. STM measurements were performed *in-situ* under UHV vacuum conditions at room temperature using an Omicron VT-STM apparatus, and data at 77 K were taken with an Omicron LT-STM instrument.

## 3. Pt nanowires on Ge(001)

*3.1 Structure with STM*
*a) Overview:*

In the following we first give an overview of the key properties of nanowire growth and structure, before addressing the question of metallic conduction states at the Fermi level. Pt nanowire structures can be seen in the STM images of Figure 1(a). The extent of the nanowires is usually terminated by terrace steps or defects. The distance between two nanowires is measured to be ~ 16 Å for most of them, while occasionally a spacing of ~ 24 Å is found - an observation consistent with Gurlu *et al.* [28]. At a tunneling bias of –1.4 V as in Figure 1(a), a *dimerization* along the chains is clearly visible in the topographic data. Along the wire, the corresponding repetition distance is ~ 8 Å. The structure between the nanowires was studied in earlier work by Gurlu *et al.* [21] and Schäfer *et al.* [30], by looking at sites with missing nanowires that expose the substrate. There a zigzag-like structure is observed, giving rise to a repetition period of also 8 Å. Concerning the correlation between the wires, as well as their structural integrity, in Figure 1(b) it emerges that the apparent period doubling (i.e., 8 Å periodicity) along the wires exists over long distances. Moreover, the surface unit cell geometry derived from Figure 1(a) and (b) can be identified as p(4×2).

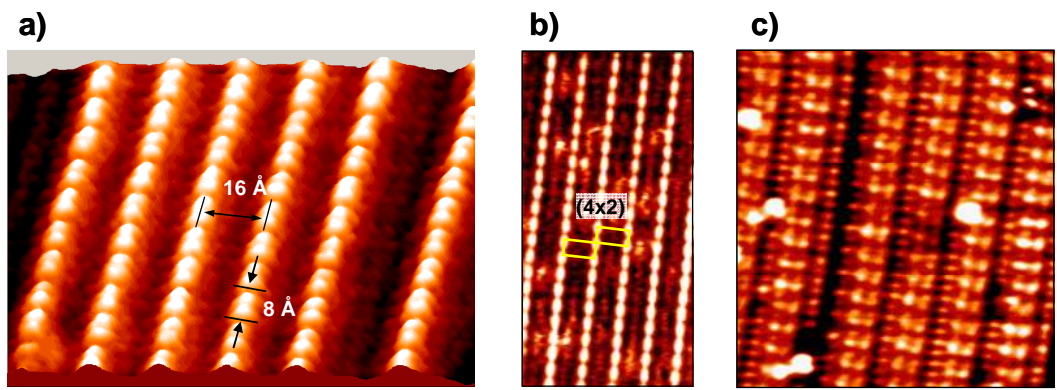

**Figure 1.** a) STM image of Pt nanowires on the Ge(001) substrate (120 Å × 90 Å, occupied states at –1.4 V). Dimerization along the nanowires (8 Å period) is observed. b) The Pt nanowires (-1.7 V) keep the registry to each other over long distances, forming a p(4×2) unit cell. c) A dimerization is also found in the unoccupied states, yet perpendicular to the chain direction (90 Å × 110 Å, +1.4 V).

Concerning the role of the *substrate geometry* in nanowire formation, it is insightful to reveal the relationship to the substrate registry by comparing to the clean Ge(001) surface prior to deposition. Here a (2×1) and a c(4×2) reconstruction coexist [36,37] at room temperature, because in this temperature range they are energetically equivalent. The atom spacing amounts to ~ 4 Å between dimers of the undistorted (2×1) phase. The repetition distance along the zigzag chains of the buckled c(4×2) phase is twice as much, ~ 8 Å. This buckling bears close resemblance to the substrate zigzag in the presence of nanowires. Despite the plain Ge surface being modified by dilute incorporation of Pt atoms [38], a direct relation to the dimer spacing of the adjacent nanowires exists, suggestive of a substrate-driven origin of the nanowire dimers. In other words, it appears that the nanowires are formed along the dimer rows of the substrate, a process which may be referred to as *template-driven* nanowire formation.

In turning from occupied to *unoccupied* states, a topography as shown in Figure 1(c) is obtained. Here the prominent structure changes to *sideways* dimers. We note in passing that a left-right asymmetry is observed in the image, indicating that the building blocks for the nanowires are asymmetrically composed. A few defects may be due to surplus physisorbed atoms, yet they do not perturb the periodicity. The superstructure along the chain direction again measures ~ 8 Å, like in Figure 1(a). From these STM measurements taken at high tunneling bias one must therefore conclude that, at least at high binding energies, a strong dimerized bonding exists, in registry with the underlying Ge surface lattice.

*b) Large area growth:*

The growth of platinum-based nanowires is strongly dependent on supplying sufficient thermal energy for the actual formation of the reconstruction. If Pt is deposited onto a cold substrate, i.e., at room temperature, a very high post-anneal temperature is required. Typically, suitable values are in the 700 °C – 800 °C range. We do observe that at these temperatures, a significant amount of Pt is lost from the surface, probably by reevaporation, or, partly, by diffusion into the bulk. Even when close to one monolayer of Pt is deposited on the plain Ge(001) surface, one observes after the anneal that the surface is not fully covered with the reconstruction.

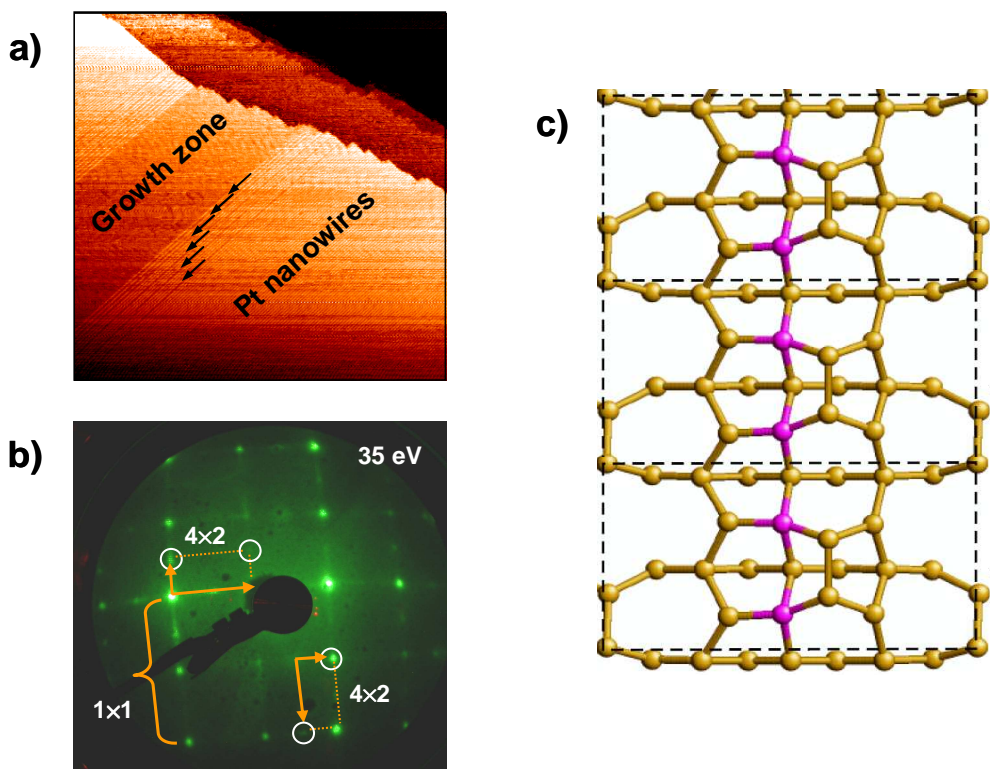

**Figure 2.** a) Large-area growth of Pt nanowires as seen in STM (2800 Å × 2800 Å, -1.7 V). The surface coverage is not fully complete, leaving small areas of initial growth stages. The nanowires spacing increases towards such zones (arrows). b) LEED image of a Pt nanowire sample (35 eV). A p(4×2) unit cell is observed, indicative of long-range order. c) Structure model of Pt nanowires derived from DFT [31]. The top ridge is formed by Ge dimers, the Pt atoms (purple) form a Pt-Ge chain that provides electron states near the Fermi level.

As an alternative approach, we have conducted studies on "hot substrate" deposition. Here the substrate is kept at elevated temperature during the electron-beam deposition of Pt. We find that a somewhat more moderate temperature than for the post-anneal can be used, ranging around ~ 500 °C. A major improvement in the surface wetting is achieved by this method, and a rather complete surface area coverage with the Pt reconstruction is obtained, while not compromising the high structural integrity of the nanowires and their low defect density. Occasionally, patches of incomplete nucleation stages of Pt are observed, if the deposition process did not supply sufficient amounts of Pt. In Figure 2(a), a growth situation is displayed where large areas of a terrace are covered by Pt, with nanowires length of at least 250 nm captured in this STM image. A fraction of the area shown, however, lacks final formation of the nanowire reconstruction. Instead, closer inspection reveals a number of atomic point defects, which may ascribed to the incomplete reaction of Pt with Ge surface atoms. In Ref. [28] this growth stage has been referred to as β-terrace. Interestingly, close to the boundary of the nanowire-covered area and the imperfect area, the spacing of the individual nanowires increases. This indicates that the amount of Pt available in this area has been marginally insufficient. Nonetheless we point out that the hot substrate process more reliably allows formation of a large area of the nanowire reconstruction.

The long-range order of the Pt nanowire reconstruction can be probed by LEED. In Figure 2(b) a LEED image of the reconstruction is shown using the hot deposition process. A p(4×2) structure can nicely be seen, in particular, with very sharp reflections. Note that there are two domains rotated by 90°, resulting from the alternating stacking sequence of the Ge substrate (which exposes both terrace types due to imperfect polishing). Such LEED pattern is an indication that the nanowires are aligned in registry over large distances. This observation is consistent with the STM images in Figs. 1(a) and (b) where a p(4×2) unit cell is identified, as well as a rather perfect structural interchain correlation.

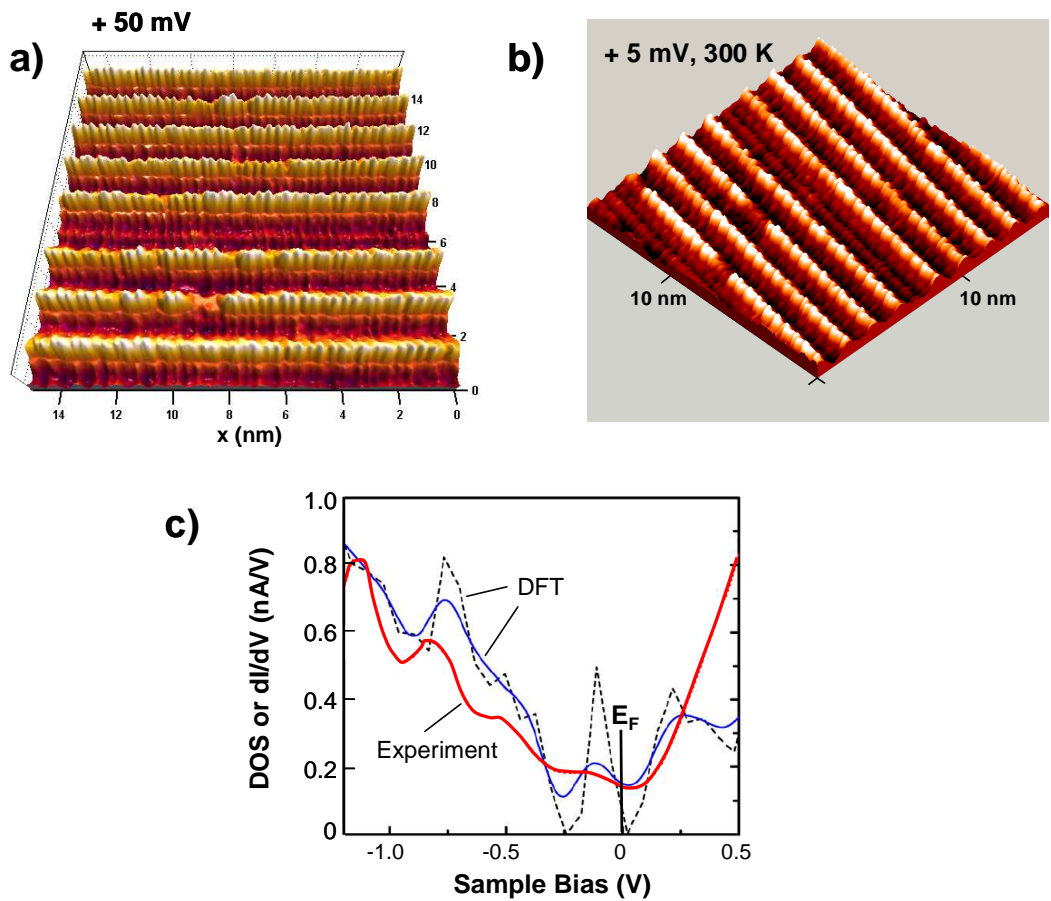

**Figure 3.** a) STM image of Pt nanowires close to the Fermi level. At +50 mV (160× 160 Å, 100pA), the dimerization is virtually absent. The image shows pronounced monomer spacing of ~ 4 Å, rather than dimers. b) Very close to the Fermi level (+5 mV), at room temperature a continuous 1D charge distribution is observed (160× 160 Å, 50pA). c) Differential tunneling conductivity at 300 K, and DFT data from the TDC model, broadened 10 meV (dashed line) and 100 meV (solid blue line).

*c) Structural model:*

An atomic structural model has meanwhile been derived [31,39], using density functional theory (DFT). A coverage of ¼ ML as published in [28] has been assumed. As a test of the validity of the structural models under inspection, simulated STM images based on the local density of states can be generated for comparison with experimental STM data. Calculations have been performed using the VASP code [39]. The DFT structure model is shown in Figure 2(c). As a key feature, the *dimers* in chain direction (observed in STM at high negative bias) are explained in this model as apparent Ge dimer units that form the top ridge of the nanowire. These two Ge atoms belong in fact to a Ge tetram-er, a structural unit which has been identified earlier on in Ge related reconstructions [40]. This combination of structural units hence lends the structure its name "tetramer-dimer-chain" (TDC) model.

In this model, the platinum is located on one side of the chain, e.g., on the left side in Figure 2(c). Moreover, the Pt atoms are connected via Ge atoms. From a study of the local orbitals, it emerges that the states near the Fermi level - which are responsible for the conduction properties - are located on this alternating Pt-Ge chain. Moreover, the orbitals here are of significant d-character [31,39]. The model can be used to predict STM images at various bias values. This has been done in Ref. [31], and one finds that the TDC model reproduces the experimental STM results very well, including low bias values, such as in the range of 0.1 eV. It also nicely shows the change from longitudinal dimer-like appearance to sideways oriented dimer-like intensity (in terms of electronic LDOS for a given energy window) when the bias is varied to detect occupied or empty states, respectively.

*3.2 Low-energy properties*

The low-energy properties of the nanowires deserve focused attention, in particular concerning the question of potential metallic character. The STM images change significantly when going to lower bias. In the occupied states closer to the Fermi level, the dimerization is largely suppressed, as in Figure 3(a). A charge density along the wires which is continuously connected can be observed close to $E_F$, e.g., if the bias amounts to V = + 50 mV as in Figure 3(a). At T = 300 K the thermal broadening of the Fermi distribution amounts to 4 kT ≈ 100 meV, thus this dataset is rather close to the conduction electron states. The variation in the charge distribution along the nanowires has become weak, and the periodicity of the remaining charge modulation is that of a *monomer* distance of ~ 4Å. We add that the detailed structure observed in STM, especially at low bias, does depend on the tunneling current [30]. If this current is not adequately reduced along with the bias, the dimer features, ascribed to the backbonds, will reemerge. In stark contrast to Figure 1(a) at high bias, the overall impression of the nanowire ensemble is here that they represent a spatially rather uniform density of states along the 1D direction, without interruption of the charge cloud between atoms.

With respect to the question of metallic behavior, we have imaged the wires at lowest voltages. The result for +5 meV is shown in Figure 3(b). This image still does also not show a prominent indication of nanowire dimerization. The point here is that the structure of the nanowires can be imaged consistently down to these exceptionally low voltages. They still exhibit a highly continuous local density of states along the 1D direction, which we take as one indication that they are *metallic* conductors. This refers to the nanowires itself and is to be distinguished from quantum well states between nanowires as in Ref. [29].

With respect to the conduction properties, tunneling current spectroscopy has also been performed [30]. At room temperature (300 K), a finite conductivity at zero bias is found, as in Figure 3(c). A marginal conductivity was also reported in [28]. From DFT and the structural model presented above one may obtain the corresponding curve, i.e., the local density of states (LDOS) integrated over the unit cell [31]. The experimental tunneling conductivity is a reflection of this DOS. DFT at T = 0 K predicts a DOS which closely follows the experimental STS data [30] and its bias-dependent features. However, it predicts a dip in the DOS at the Fermi level, corresponding to a zero energy gap resulting from a semi-metal, as can be seen in Figure 3(c). At 300 K, the features in the curve will be thermally broadened (including the broadening of the Fermi distribution in the tunneling tip). For suitable broadening, the DFT data correspond well to the STS data at 300 K [31], consistent with a small but finite conductivity at room temperature.

## 4. Au-induced chains on Ge(001)

*4.1 Structure with STM*
*a) Overview:*

The self-organized growth of the Au-induced nanowire reconstruction has excellent wetting capabilities concerning the Ge(001) substrate, much more so than for Pt. This is further enhanced by using a hot substrate during growth. These Au nanowires accordingly cover the whole substrate area, as in the STM image of Figure 4(a). The reconstruction runs up to occasional terrace edges of the Ge(001) wafer. Thus there is no principal length limitation to the nanowires, which easily extend from several 1000 Å into the μm-range. If a terrace step occurs, the orientation of the Ge(001) dimer rows as well as the Au nanowires rotates by 90°. The nanowires are aligned in parallel with very even charge density in 1D direction. They follow the direction of the underlying dimer rows of the (001) surface (as in the Pt case), for which reason this situation can again be viewed as template-driven self-organization. The lateral chain spacing amounts to 16 Å, corresponding to 4× the Ge atom distance of the Ge(001) surface. It is striking how well the chain ridges are separated, with a deep minimum in-between. Individual atoms are not identified for most tunneling conditions, indicative of a strong 1D delocalization of the electron states.

These Au chains exhibit excellent long-range order. It can be probed on an extended sample area by LEED, as in the data of Figure 4(b). One observes a well-defined c(8×2) reconstruction, allowing for two surface domains. These long-range ordered c(8×2) nanowires were first reported by Schäfer *et al.* [41], and the results strongly depend on an accurate coverage (slightly in excess of 0.5 ML) and on using a hot substrate deposition process. In an earlier work, growth of Au was reported by Wang *et al.* [32], however, they arrived at a (4×2) LEED pattern. There the authors have been using an unusually high Au coverage of 1.5 ML, for which reason probably a different surface phase was obtained. In that early study, the STM data also showed lots of defects. In the recent work on the c(8×2) phase, this growth regime has been avoided, aiming at a lower coverage and at a perfected long-range order.

*b) Structural features:*

The STM images deserve some closer attention concerning the structural features and the unit cell. The STM image of Figure 5(a) with bias -1.0 V reveals a weak vertical undulation on the scale of 0.3 Å, as shown in the profile along the wire in Figure 5(b). It reflects the lattice periodicity of 2a = 8

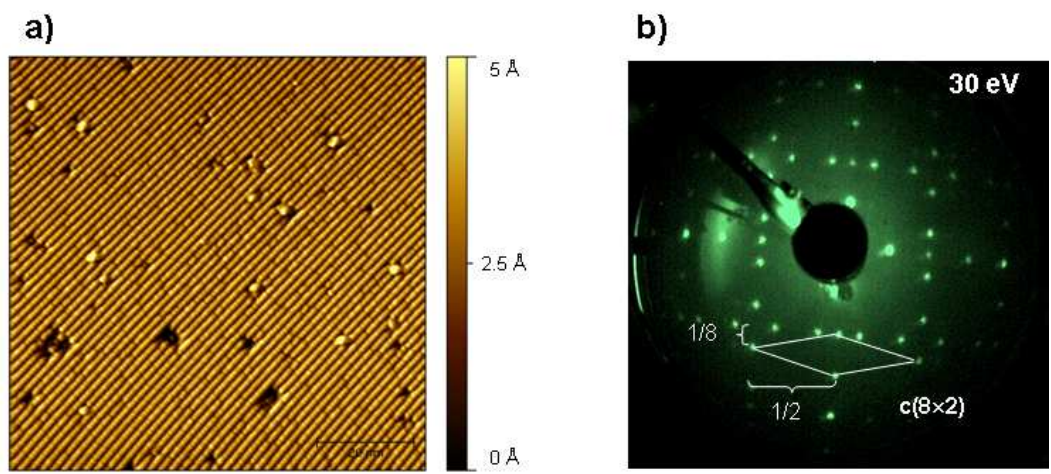

**Figure 4.** a) STM image with overview of Au/Ge(001) nanowires (840 Å × 840 Å, +0.8 V). The nanowire spacing is 16 Å, with even intensity in 1D direction. b) LEED image at 30 eV, showing a sharp c(8×2) structure (with dual domains), corresponding to excellent long-range order.

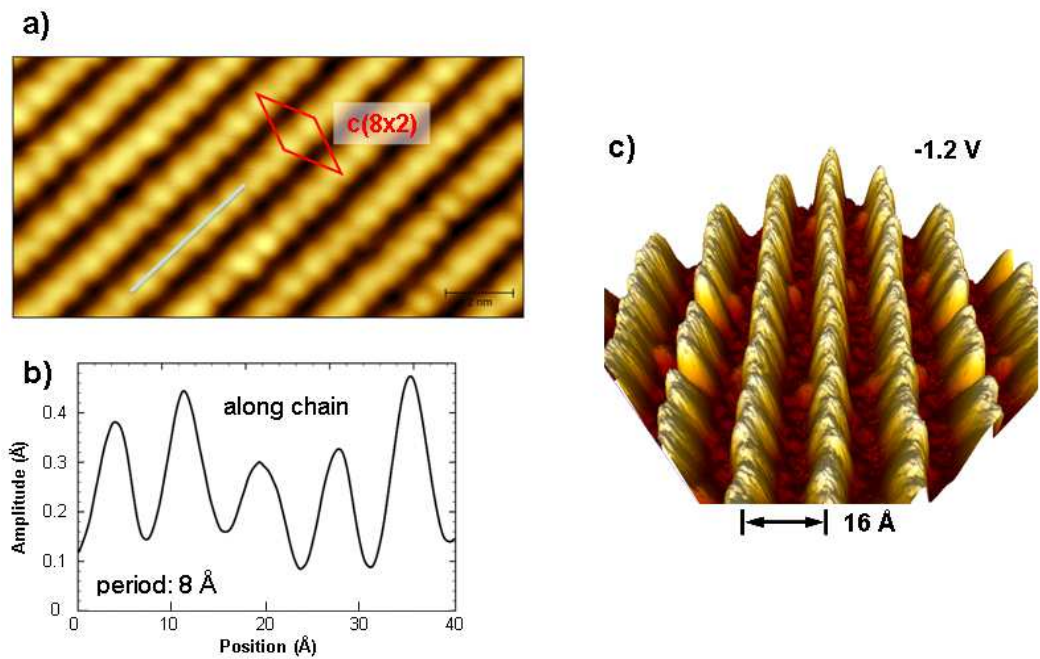

**Figure 5.** a) Au chain STM image at -1.0 V (0.4 nA, T = 77 K) with weak modulation corresponding to c(8×2) unit cell. b) Length profiles confirm 8 Å periodicity, and a weak modulation of ~ 0.3 Å. c) Topographic view (100 Å × 100 Å) of the Au nanowires (-1.2 V, 0.4 nA). A significant spacing between the nanowires can be seen. The nanowires are atomically narrow, with slight zigzag along the ridge.

Å along the chain. Neighboring chains are shifted by half this period length, which accounts for the centered c(8×2) unit cell. The difficulty to observe this periodicity in STM can be understood from the delocalized charge cloud which does not allow to see atomic structure details on the wire or from its deeper lying construction units.

For the lateral electron confinement, STM yields an astoundingly sharp top contour of the nanowire, as in Figure 5(c) with a rather large height of typically ~ 1-2 Å, depending on tunneling conditions. One sees very nicely how the nanowire ridges are sharply elevated above the Ge substrate, and how they are separated from each other. Moreover, along the 1D direction, a marginal zigzag contour exists on top of the wires. However, in comparing this to the rather well known dimer buckling on the plain Ge(001) surface, the weak left-right buckling of the nanowires observed in STM appears to be less pronounced. Apart from this undulation, one observes a rather continuous linear extent of the charge distribution.

With respect to the lateral extent of the nanowire ridge, the resolution broadening by the STM tip needs to be considered, which results from the tip diameter and the tunneling gap. One has thus to consider a broadening of several angstroms, as was, e.g., demonstrated for Na atoms [42]. For details of this analysis, the reader is referred to [41]. Accurate profile analysis of the Au nanowires yields an apparent width of ~ 7 Å including these broadening mechanisms. This interestingly compares well to (and is actually slightly less than) the profile width for a *single Au atom* on an insulating alumina substrate [43]. This suggests that the top charge cloud is confined laterally comparable to the size of a *single* atom. We emphasize that STM detects valence orbitals rather than atom positions, nor is it sensitive to the chemical element species. In particular, the observed charge density distribution can also reflect the spill-out from deeper layers or a bond between two atoms. However, these considerations do not affect the observation that the lateral charge confinement has obviously reached the atomic limit.

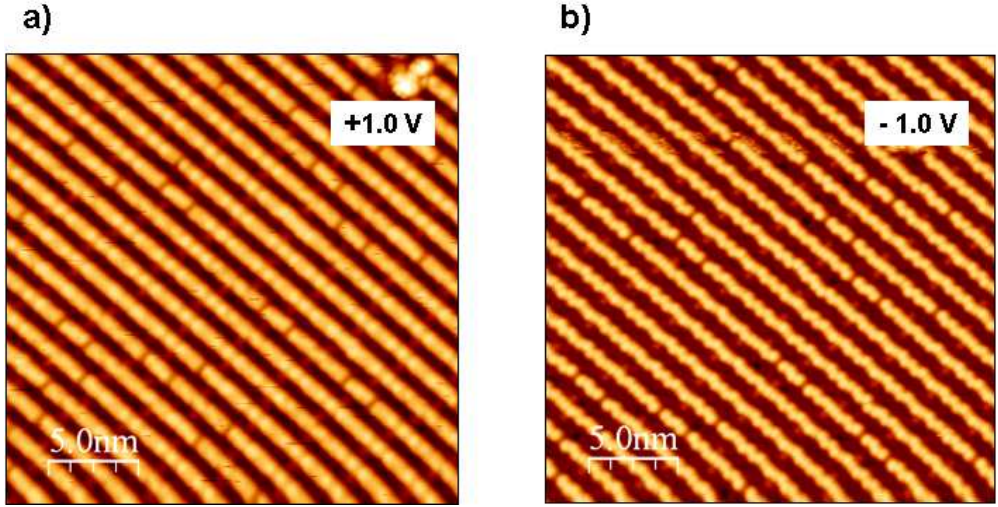

**Figure 6.** a) STM image at +1.0 V (250 Å × 250 Å, T = 77 K) of Au nanowires, providing evidence for a strict spatial separation. b) Same area at -1.0 V. The structure looks largely unchanged (allowing for a weak zigzag modulation), without any cross-bonds between the wires.

In closely inspecting the architecture of the 1D surface reconstruction, we have recorded low-noise STM data as in Figure 6. A sharp nanowire shape is consistently obtained when the bias is varied from positive to negative, as in Figure 6(a) or (b), respectively. It is very striking that the images hardly change when the bias is varied so drastically, including sign reversal. Moreover, and very importantly, in neither image, Figure 6(a) and (b), does one observe cross-links between the nanowires, or any other type of two-dimensional pattern. This is very different from the STM measurements of the Pt nanowires on Ge(001) [30], where variation of bias does unveil sideways bonds, as discussed in the preceding section. In the case of the Au-induced chains, no such additional bonds emerge, and the nanowires appear unusually homogeneous and structureless. The charge density reaches a peak along the nanowire, falling off to both sides. Even the detected nanowire width is essentially bias-independent. It thus emerges as a key fact that the deeply modulated ridges reflect true topography, i.e., the architecture of the nanowires.

*4.2 Low-energy properties*

In addressing the electronic states close to the Fermi level, the delocalized charge in 1D direction persists even for low bias values, such as -0.1 V and + 0.1 V, as shown in Figure 7(a). A sharp nanowire contour is consistently obtained even under these tunneling conditions. The shape at a bias magnitude of 0.1 V is effectively independent of polarity. We conclude that the consistently even chain appearance requires 1D electron levels with density of states both below and above the Fermi level $E_F$. The only viable explanation is that a rather delocalized 1D electron system resides on top of the nanowires, whose charge cloud dominates the tunneling signal.

Exact information on the local DOS around the Fermi level is obtained from tunneling spectroscopy. The differential conductivity obtained at room temperature on the nanowires is shown in Figure 7(b) (averaged along the chains). For the Au chains, a substantial tunneling conductivity is detected around $E_F$. This is evidence for a metallic character. Between the nanowires, the tunneling conductivity declines. Such measurements require to position the tip accurately and to avoid signal from the sloping side of the nanowires. This inhomogeneity demonstrates how well the charge is confined to the chain.

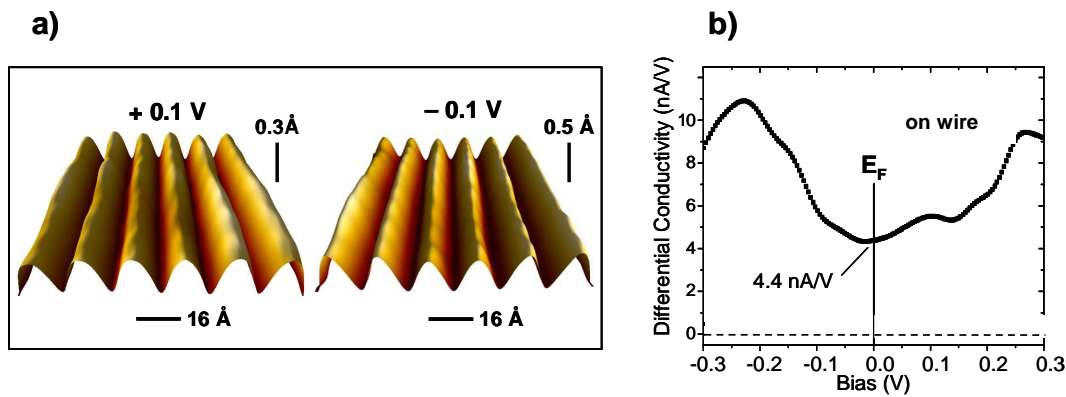

**Figure 7.** a) Au nanowire STM (100 Å × 100 Å) at +0.1 V and -0.1 V. Even at such low bias, the data indicate the characteristic sharp ridge, as well as 1D charge delocalization of the states near $E_F$. b) Differential conductivity on nanowire top at 300 K. Significant metallic conductivity is observed.

Here we add that metallic bands have also been found independently in angle-resolved photoemission (ARPES) measurements [41]. A shallow band is seen with a minimum of ~ 130 meV below $E_F$. It carries very roughly the shape of an upward parabola extending through $E_F$. Such *metallic* band is not known from bulk Ge nor from the clean Ge surface [44], and must be ascribed to the Au-induced surface reconstruction.

## 5. Discussion and comparison

In reviewing the major findings about nanowires induced by the noble metals Pt and Au on Ge(001), there are a number of key characteristics that are very different from the reconstructions known on, e.g., Si(111). With respect to the architecture of the Pt nanowires on Ge(001), our experimental results suggest that dimer units are the building blocks of the nanowire embankment. A dimerization period of ~ 8 Å in 1D direction is readily apparent in most STM images, however, as seen at varied tunneling bias, also sideways bond orientations do occur. The longitudinal dimer structure has also been noted by Gurlu *et al.* [28]. The DFT calculation and the structure model derived from it [31,39] now provides an explanation within the tetramer-dimer-chain model. It does unveil that the most prominent dimer structure is Ge-derived, a finding that has been confirmed in an independent DFT study for varied coverages [45]. Also, there is a rather complex backbonding to the substrate. The latter does involve as key characteristic an alternating sequence of Pt and Ge atoms in chain direction on the side of the nanowire.

The electronic conduction properties are determined very close to the Fermi level $E_F$, were the experiment finds non-dimerized charge clouds that relatively evenly spread out along the wire. The sharp monomer STM data observable in a window from approximately $E_F - 0.2$ eV to $E_F + 0.2$ eV are in fact suggestive of spatially well-defined and evenly spaced orbitals that stick out of the chains. From DFT one may actually derive the orbital nature of the LDOS at various binding energies [39]. Most of the states are Ge-derived, with s- and p- orbitals contributing to the intense LDOS at high binding energies of the order of eV energies. The Pt atoms are adding their 5d and 6s orbitals. Especially, the d-orbitals are donating charge into the Pt-Ge chain, which contributes to the lower binding energy states, and thereby to the conduction states at finite temperature. The local charge distribution is reflected in simulated STM images calculated by Stekolnikov *et al.* [31,39] that show how charge resides at the Pt atoms at low bias. The actual charge distribution is in fact rather subtle, and does include LDOS at various neighboring Ge atoms (including the top ridge atoms). The agreement of the predicted STM images based on the TDC model with the experimental data is good throughout for a large range of bias values.

The resulting overall atomic structure of the nanowires formed by Pt must hence be called rather special. Most notably, the top dimer ridge contains dimers *along* the chain direction, in contrast to the plain Ge(001) surface, where the dimers are oriented perpendicular to the chain direction. Moreover, the plain Ge(001) surface is significantly reconstructed, with hardly any bonding element being reminiscent of the underlying surface. This may partly explain why the surface formation temperature (i.e., activation temperature) is so exceptionally high. In fact, during the high temperature regime required for self-organized growth (~ 500 – 700 °C), one already observes significant evaporation of Pt from the semiconductor surface, associated with a reduction in Pt coverage. In turn, once formed, this Pt nanowire reconstruction is extremely stable over time, and upon subsequent annealing treatments.

In turning to the Au-induced chains, the information on atomic structure is much less straightforward, as the STM images do not reveal individual atom positions, and show a smeared-out charge distribution instead. This holds for a large range of bias values irrespective of polarity. Also, currently no DFT calculation exists that would resolve the structural model. Thus far, an extreme charge confinement to the atomic scale has been found in the STM data, while at the same time, from tunneling spectroscopy, the nanowires seem to be solidly metallic. It has been suggested in a STM study [46] that dimers on top of the nanowires are a possible ingredient in such atomic structure. However, more work seems to be needed to clarify this issue, especially including DFT modeling to judge on the energetic stability of the structural ingredients. DFT studies on various Au-Ge bonding situations suggest that the energy differences may in fact be rather subtle [47], providing a challenge for producing a structural model.

One may thus ask the question whether the electronic states in these Au chains are a potential realization of a 1D electron liquid, for which quantum theory predicts exotic properties. Specifically, the Fermi liquid paradigm is predicted to be no longer valid in 1D, and it is generally assumed that the electrons form a Luttinger liquid [2,3,48,49]. Such electron liquid is predicted to show power-law scaling of the spectral weight around $E_F$, with a depression down to zero weight (observable ideally at zero temperature, and with thermal melting at finite temperature).

There are some indications which suggest that the Au nanowires might potentially be compatible with such picture, albeit that detailed investigations still need to be made. On one hand, there is no evidence of long-range CDW ordering at 80 K in LEED. This is consistent with a strict 1D regime in which a CDW phase will be suppressed due to fluctuations. Secondly, the spectral weight in ARPES at 80 K is found to be significantly reduced towards $E_F$ in a ~ 30 meV region [41]. This points at the fascinating question whether the system exhibits the theoretically predicted properties at low temperature. A suitable method to detect the predicted power-law spectra is to conduct low-temperature scanning tunneling spectroscopy (STS) [50]. Equivalently, one may study the spectral weight in angle-resolved photoemission.

In comparing the Au chains on Ge(001) to other well known, truly metallic nanowire reconstructions, there are significant differences. Most notably, the previous reconstructions seem to provide not as good a structural separation between the nanowires. The well-studied In-nanowires on Si(111) [9] consist of four metal atoms embedded between Si rows, and exhibit a CDW below ~130 K. In Au/Si(557) and Au/Si(553) [10,19], bonding to and between substrate atoms seems to govern the electronic properties [22,51]. These systems show CDW condensation already slightly below room temperature [10]. It is important to note that a high CDW condensation temperature requires significant interchain coupling, because otherwise thermal 1D fluctuations suppress the condensation of the ground state at finite temperature. Therefore, a high CDW critical temperature does necessarily imply a significant lateral coupling. In stark contrast, the present Au chains are spatially exceptionally well separated (16 Å), which is far beyond orbital overlap, and confined to an atomic-scale metallic path.

Their conduction is quite apparently decoupled from the Ge(001) substrate, which is an insulator both in the bulk as well as for the plain Ge(001) surface [44].

## 6. Conclusion

In summary, a new class of nanowires is established on the Ge(001) surface, formed by the noble metals Pt and Au. In a comprehensive study of both nanowire types, tunneling images over a wide range of unoccupied and occupied states were obtained, unveiling an unprecedented spatial separation of the chains.

In the Pt nanowires, although dimer elements exist with longitudinal orientations, no obvious connection to a CDW can be made. Instead, closer to the Fermi level, the dimerization appears lifted and nanowire atoms with monomer spacing become visible. Their metallic conductivity arises from an independent electron band, partially related to the Pt atoms which are not involved in the dimer formation of the wire top. A low density of states near the Fermi level is found in DFT, in fact associated with a semi-metal behavior, which nonetheless in the experiment at room temperature leads to a finite metallic conductivity.

The Au atom chains on Ge(001) grow in similarly well separated manner, albeit that their structural model still remains unresolved. They embody a unique 1D electron system with unprecedented properties. It comprises an almost fully delocalized charge distribution in chain direction found in STM. At the same time, the confinement is laterally so narrow that the atomic limit has been reached.

Both systems do provide new stimulus for studying unconventional physics in a 1D electron liquid. Specifically, there are important questions to address by low-temperature spectroscopy. Further insight might be provided by investigations of a possible CDW instability of these electrons at very low temperature. In the Pt chains, based on the DFT results [31,39] this is not expected due to a lack of a Fermi surface nesting condition, and in fact the DFT study suggests even a semi-metallic situation. However, there have been indications in a low-temperature STM study that point at a distortion along the chains [52], the origin of which awaits further exploration and might relate to structural contributions. For the Au chains, experimentally no indication has been found yet at intermediate temperatures, while helium cooling studies are still underway. The same holds for studies of the spectral function, which need to be performed by STS and ARPES, and which may unravel unusual phenomena.

More generally, these experiments with successful incorporation of different noble metal atoms into the Ge(001) surface demonstrate impressively how the properties of atomic self-organized structures can be *tailored* to yield certain desired properties. In this particular case, by changing the atom species from Pt to Au, the orbital character of the active valence electrons has been changed (from d-like to more s-like). Consequently, a conceivably different wire architecture has been realized, as well as notably different electronic properties (dimerized bonds versus continuous 1D charge density in Pt or Au nanowires, respectively). Moreover, one may expect that, by addition of extra atoms that act, e.g., as dopants, one may tune the charge content of the chain, and thereby its electronic properties. This demonstrates vividly how *atomic control* is possible in these nanostructures at the ultimate lower size limit.

Independent of the low-energy electron states, as an outlook one may also think of using these nanowires as a template in adsorption studies. The preformed nanowire ridges may, e.g., serve as alignment aid (i.e., a *template*) for the adsorption of other atom species that do not form chain reconstructions by themselves. It may equally well be used as a template for adsorption of molecules, which by this means can be artificially aligned. Such *template-induced structuring* may lead to new kinds of long-range order, and thereby specific interactions between the molecules. Finally, one may also take advantage of the catalytic activities of the noble metal atoms, especially Pt. By integrating Pt in a stable manner onto the surface, as is the case for the nanowires, one may in fact have a catalyst with a high degree of temperature stability. Experimental demonstration of catalytic activity on the Pt nanowires would thus be a fascinating chemistry experiment on the atomic scale.

**Acknowledgement**

We are grateful for discussion with F. Himpsel, J. McChesney and H. Pfnür. This work was supported by the Deutsche Forschungsgemeinschaft (grant Scha1510/2-1).

J. Schäfer *et al.*